\documentclass[runningheads]{llncs}
\usepackage[T1]{fontenc}
\usepackage{graphicx}
\usepackage{multirow}
\usepackage{makecell}
\usepackage[caption=false]{subfig}
\usepackage[table]{xcolor}
\newcolumntype{g}{>{\columncolor{lightgray}} c}
\usepackage{cite}

\usepackage[inline=true, margin=false]{fixme}
\usepackage{hyperref}

\usepackage{color}

\begin{document}
\title{Temporal Motif Participation Profiles for Analyzing Node Similarity in Temporal Networks}
\titlerunning{Temporal Motif Participation Profiles for Analyzing Node Similarity}
\author{Maxwell C. Lee \and
Kevin S. Xu}
\authorrunning{M.~C. Lee and K.~S. Xu}
\institute{Case Western Reserve University\\
10900 Euclid Avenue, Cleveland, OH 44106-7071, USA\\
\email{\{mcl154,ksx2\}@case.edu}}
\maketitle              % typeset the header of the contribution
\begin{abstract}
Temporal networks consisting of timestamped interactions between a set of nodes provide a useful representation for analyzing complex networked systems that evolve over time. 
Beyond pairwise interactions between nodes, temporal motifs capture patterns of higher-order interactions such as directed triangles over short time periods. 
We propose \emph{temporal motif participation profiles (TMPPs)} to capture the behavior of nodes in temporal motifs. 
Two nodes with similar TMPPs take similar \emph{positions} within temporal motifs, possibly with different nodes. 
TMPPs serve as unsupervised embeddings for nodes in temporal networks that are directly interpretable, as each entry denotes the frequency at which a node participates in a particular position in a specific temporal motif.
We demonstrate that clustering TMPPs reveals groups of nodes with similar \emph{roles} in a temporal network through simulation experiments and a case study on a network of militarized interstate disputes.

\keywords{Temporal motif \and Node embedding \and Role-based embedding \and Dynamic network \and Militarized interstate dispute.}
\end{abstract}
\section{Introduction}
Temporal or dynamic networks are a natural representation for modeling time-varying complex systems such as social interaction networks, where a node denotes a person or group of people and an edge denotes some type of interaction between two nodes. 
Higher-order structures involving more than two nodes are often of great interest when analyzing networks \cite{benson2016higher}, including temporal networks.

\begin{figure}[t]
    \centering
    \subfloat[All possible 3-edge temporal motifs]{
        \includegraphics[width=0.63\linewidth]{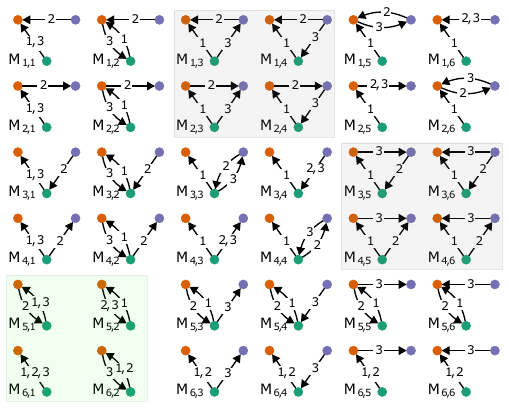}
        \label{fig:all_temporal_motifs}
    }
    \hfill
    \begin{minipage}[b]{0.32\linewidth}
    \subfloat[Temporal motif counts]{
        \includegraphics[width=\linewidth]{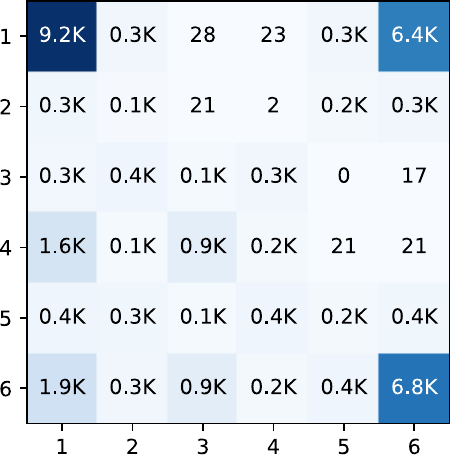}
        \label{fig:mids_total_motif_counts}
    }
    \hfill
    \subfloat[Temporal motif positions]{
        \includegraphics[width=\linewidth]{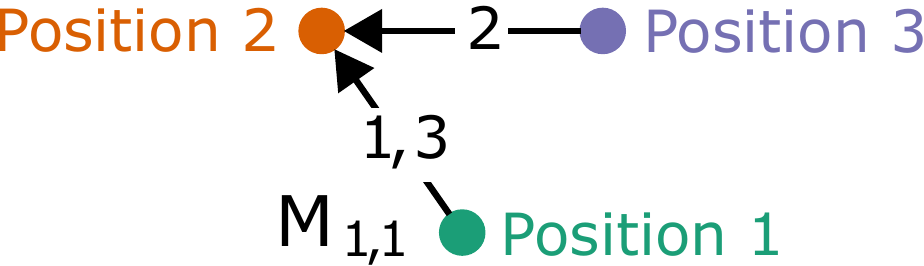}
        \label{fig:motif_positions}
    }
    \end{minipage}
    \caption[]{\subref{fig:all_temporal_motifs} All possible 3-edge temporal motifs involving 2 or 3 nodes. The edges are labeled in the order in which they appear. The green box in the bottom left denotes motifs with 2 nodes, and the grey boxes denote triangle motifs. Figure credit: \cite{paranjape2017motifs}. 
    \subref{fig:mids_total_motif_counts} Occurrence frequencies of all temporal motifs in a militarized interstate dispute network, arranged to match the motif ordering in \subref{fig:all_temporal_motifs}. Figure credit: \cite{do2022analyzing}.
    \subref{fig:motif_positions} Node positions in a temporal motif used in this paper, shown on motif $M_{1,1}$. The green node always denotes position 1, the red node always denotes position 2, and the blue node always denotes position 3. There is no position 3 for the 2-node motifs.}
    \label{fig:intro_temporal_motifs}
\end{figure}

A common way of studying higher-order relations in temporal networks is to investigate the presence of different \emph{temporal motifs}. 
Motifs are small subgraph patterns in networks \cite{milo2002network}. 
Temporal motifs further specify the order in which edges in a motif must appear, along with a maximum time duration \cite{paranjape2017motifs}. 
All of the possible 3-edge temporal motifs involving 2 or 3 nodes (assuming directed edges) are shown in Figure \ref{fig:all_temporal_motifs}.
Temporal motifs can be used to model a variety of phenomena in social networks, including escalations between countries in militarized interstate dispute (MID) networks \cite{do2022analyzing}. 

Most prior research on temporal motifs has focused on counting the number of occurrences of a particular motif in a temporal network \cite{paranjape2017motifs,do2022analyzing,liu2019sampling,porter2022analytical}. 
The occurrence frequencies (counts) of the different motifs yield insights into the behavior of the network. 
For example, the counts in Fig.~\ref{fig:mids_total_motif_counts} show that motifs $M_{1,1}$, $M_{1,6}$, and $M_{6,6}$
% , denoting 3 edges from 2 nodes to 1 node, 
occur most frequently in MID networks, while the motifs denoting temporal triangles occur least frequently.
This approach provides for a global analysis of temporal motif patterns but does not characterize how different nodes engage in different temporal motifs. 

\paragraph{Our Contribution}
Our main contribution in this paper is the notion of a \emph{temporal motif participation profile (TMPP)}, which captures how a particular node participates in different temporal motifs and in what \emph{node positions}. 
We illustrate node positions in a temporal motif in Fig.~\ref{fig:motif_positions}. 
In this motif, positions 1 and 3 are senders, while position 2 is a receiver. 
This asymmetry of node positions is present in all motifs in directed temporal networks due to both the directed nature and  temporal ordering of edges. 
The nodes in positions 1 and 3 take on very different roles from the node in position 2. 
Thus, considering how frequently a node participates in each position of each temporal motif allows us to characterize the role of the node in the temporal network.

TMPPs can be viewed as unsupervised and human-interpretable node embeddings, where each dimension of a node's TMPP maps to a particular node position in a specific temporal motif. 
This allows TMPPs to be clustered to identify groups of nodes with similar TMPPs and then visualized to understand the meaning of the cluster. 
We find that inclusion of positions in temporal motifs results in clusters of nodes that take \emph{similar roles} in a temporal network. 
On the other hand, excluding positions
results in clusters of nodes that simply participate in a lot of the same edges that form a temporal motif rather than identifying any structural roles in the network.
We demonstrate the utility of TMPPs in a case study on a MID network, where 
we identify clusters of countries that engage in similar \emph{patterns of incidents} in MIDs, but not necessarily in the same disputes, demonstrating their role similarity.

\section{Background}

\subsection{Motifs in Temporal Networks}
A temporal or dynamic network differs from a static network in that edges between nodes occur only at specified times. 
A temporal edge between nodes $u$ and $v$ is specified as a triplet $(u, v, t)$ where $t$ denotes the time at which an edge occurs. 
If $t$ only takes on a small set of possible values (e.g., $1, 2, 3, \ldots$) in a temporal network, then it is usually referred to as a discrete-time network, while a temporal network with a wide variety of values for $t$ (e.g., at the level of one day in a dataset that spans many years) is often referred to as a continuous-time network\footnote{Time is not truly continuous in any dataset, because it is collected with finite resolution; however, when the time resolution is quite fine grained with respect to the total length of the dataset, modeling time as continuous is a reasonable choice.}. 
We consider directed continuous-time networks in this paper, so that a temporal edge $(u,v,t)$ denotes an edge from node $u$ (the sender) to node $v$ (the receiver) at time $t$.

One way of analyzing networks is to consider the frequencies of at which small subgraphs called \emph{motifs} occur in the network \cite{milo2002network}. 
A (non-temporal) motif in a directed network is made up of three or more edges between two or more nodes. 
There are several different definitions for temporal motifs; see the survey by Sar{\i}y{\"u}ce \cite{sariyuce2025powerful}. 
We use the definition by Paranjape et al.~\cite{paranjape2017motifs}: 
a temporal motif in a directed temporal network is a sequence of three or more edges between two or more nodes, where the times of the edges must occur \emph{sequentially, but not necessarily consecutively,} within some specified period of time.

In this paper, we consider all temporal motifs with 3 edges between 2 or 3 nodes, as shown in Fig.~\ref{fig:all_temporal_motifs}. 
Our proposed approach is not specific to the type of temporal motifs and can be extended to motifs with larger numbers of nodes or edges or specific types of motifs. 
It is also not specific to the definition of temporal motif and can be used with the other definitions of temporal motifs described by Sar{\i}y{\"u}ce \cite{sariyuce2025powerful}.

\subsection{Related Work}

\paragraph{Role-based Node Embeddings}
Our proposed TMPPs capture how nodes participate in different positions in temporal motifs, which correspond to different roles, using a vector representation for each node. 
This approach shares some similarity with role-based node embedding approaches such as Role2Vec \cite{ahmed2020role}, among others \cite{rossi2020proximity}. 
Role-based embeddings are typically learned for some type of supervised learning task, including node classification. 
Such embeddings are not human interpretable, as they are simply vector representations for nodes that optimize accuracy for the supervised learning task.
On the other hand, our TMPPs are computed in an unsupervised manner and result in human-interpretable vector representations, as we describe in Section \ref{sec:tmpp}. 
While they can be used also for supervised learning tasks, their primary benefit is to allow for analysis of temporal networks, not to optimize prediction accuracy.

\paragraph{Temporal Motif-based Models}
Another related line of research focuses on models for generating temporal networks with temporal motifs. 
Porter et al.~\cite{porter2022analytical} propose the Temporal Activity State Block Model (TASBM), which divides nodes into groups or blocks and generates continuous-time networks with timestamped edges. 
This model provides analytical expressions for the expected counts and variances of different temporal motifs across a network. 
Soliman et al.~\cite{soliman2022multivariate} propose the Multivariate Community Hawkes (MULCH) model, which also divides nodes into blocks and generates timestamped edges using structured multivariate Hawkes processes \cite{laub2021elements}. 
They demonstrate that the MULCH model can generate networks with a wide variety of temporal motif distributions, and these distributions can closely match those of the network that is being used to fit the model. 
Other modeling approaches including Structural Temporal Modeling
(STM) \cite{purohit2018temporal} and the Motif Transition Model (MTM) \cite{liu2019sampling} directly try to generate temporal networks using temporal motifs. 

These models typically validate their ability to generate temporal motifs by examining the total counts of motifs generated by the model and comparing them to the counts from an actual network, e.g., as shown in Fig.~\ref{fig:mids_total_motif_counts}. 
They do not consider temporal motif counts at the level of individual nodes, nor do they consider different positions in the motifs, which is the main focus of this paper.

\paragraph{Dynamic Graphlets}
Another related line of research is on dynamic graphlets \cite{hulovatyy2015exploring,aparicio2018graphlet}. 
Graphlets share some similarities with network motifs as they are both subgraphs; however, graphlets also account for node symmetries in subgraphs through automorphism orbits. 
Both \cite{hulovatyy2015exploring,aparicio2018graphlet} discuss how a node's involvement in dynamic graphlets can reveal information about its behavior, which is also the main idea behind TMPPs. 
However, we use our proposed TMPPs to analyze participation in temporal motifs with directed edges, where different node positions in a temporal motif can indicate very different behaviors.

\section{Methods}
\label{sec:methods}

\subsection{Temporal Motif Participation Profile (TMPP)}
\label{sec:tmpp}
We characterize the role of a node in a temporal network through its \emph{temporal motif participation profile (TMPP)}, which is a vector representation of its participation in different positions of different temporal motifs. 
We consider all possible 3-edge motifs involving 2 or 3 nodes.
There are 32 such motifs involving 3 nodes and 4 such motifs involving 2 nodes (in the green box in the bottom left corner of Fig.~\ref{fig:all_temporal_motifs}), resulting in $32(3) + 4(2) = 104$ possible node positions. 
For each node, we construct a 104-dimensional count vector, where each cell denotes how many times a node participates in a motif position. 

Different nodes may participate in drastically different numbers of temporal motifs; for example, Do and Xu \cite{do2022analyzing} found that the distribution of temporal motifs over nodes in MIDs is heavy tailed. 
Hence, we normalize the count vector by its sum to form the TMPP. 
The TMPP for each node is thus a 104-dimensional vector, where each entry 
is a normalized counts indicating how frequently that node participates in a particular temporal motif and in what position. 
The TMPP can be viewed as a type of role-based embedding, as nodes that take on similar positions in temporal motifs would have similar TMPPs.

\begin{figure}[tp]
    \centering
    \subfloat[Constructing temporal motifs]{
        \includegraphics[width=0.49\linewidth]{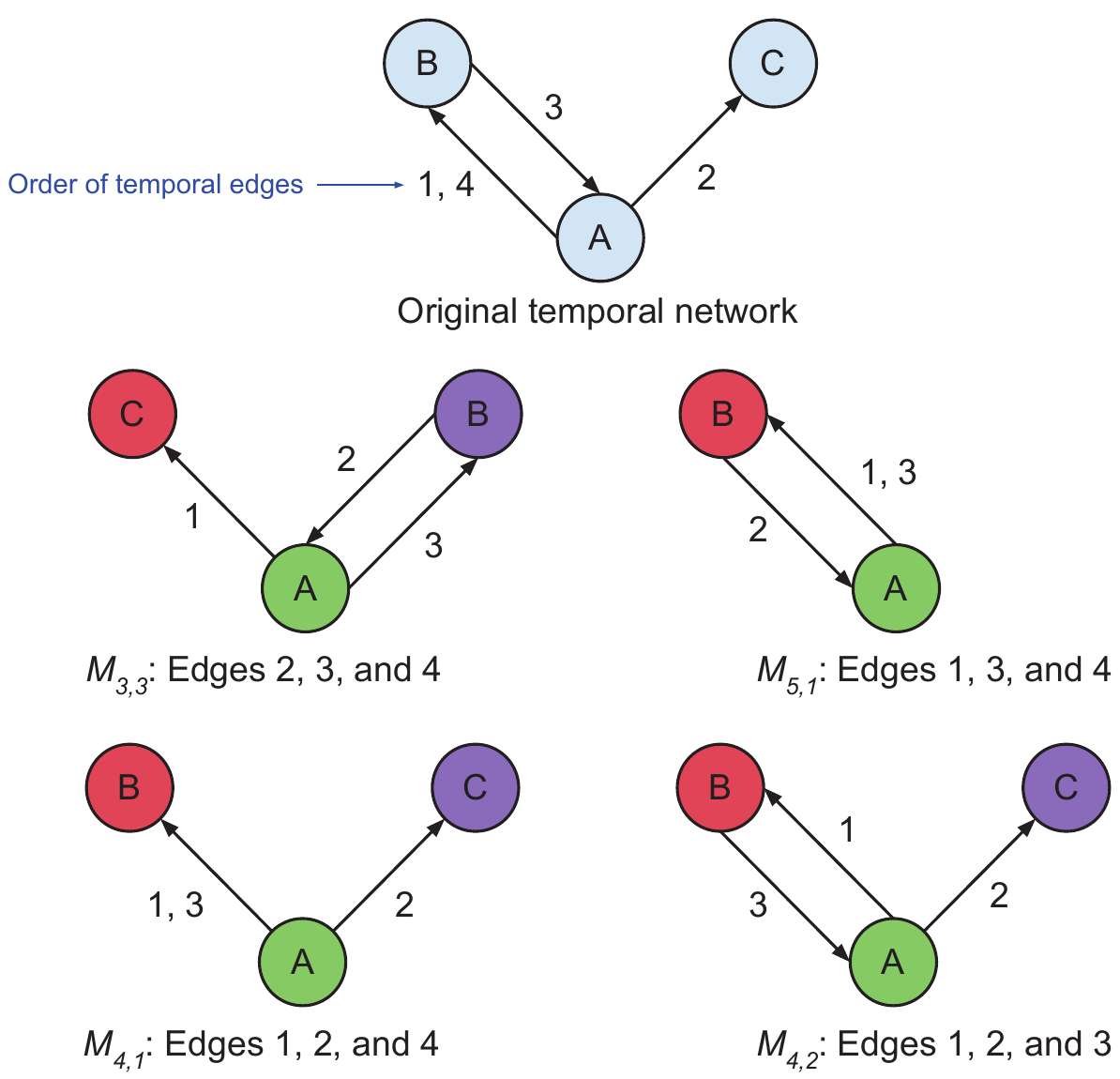}
        \label{fig:tmpp_example_graph}
    }
    \hfill
    \subfloat[Temporal motif position counts]{
        \renewcommand{\arraystretch}{1.1}
        \begin{tabular}[b]{ccgcc} 
        \hline
        Motif & Position & Node A & Node B & Node C \\
        \hline
        \multirow{3}{*}{$M_{3,3}$} 
                       & 1 & 1 & 0 & 0 \\ 
                       & 2 & 0 & 0 & 1 \\ 
                       & 3 & 0 & 1 & 0 \\ 
        \hline
        \multirow{2}{*}{$M_{5,1}$} 
                       & 1 & 1 & 0 & 0 \\ 
                       & 2 & 0 & 1 & 0 \\ 
        \hline
        \multirow{3}{*}{$M_{4,1}$}
                       & 1 & 1 & 0 & 0 \\ 
                       & 2 & 0 & 1 & 0 \\ 
                       & 3 & 0 & 0 & 1 \\ 
        \hline
        \multirow{3}{*}{$M_{4,2}$}
                       & 1 & 1 & 0 & 0 \\ 
                       & 2 & 0 & 1 & 0 \\ 
                       & 3 & 0 & 0 & 1 \\ 
        \hline
        \end{tabular}
        \label{fig:tmpp_example_counts}
    }
    \hfill
    \subfloat[Temporal motif participation profile (TMPP) vectors for all nodes]{
        \includegraphics[width=0.9\linewidth]{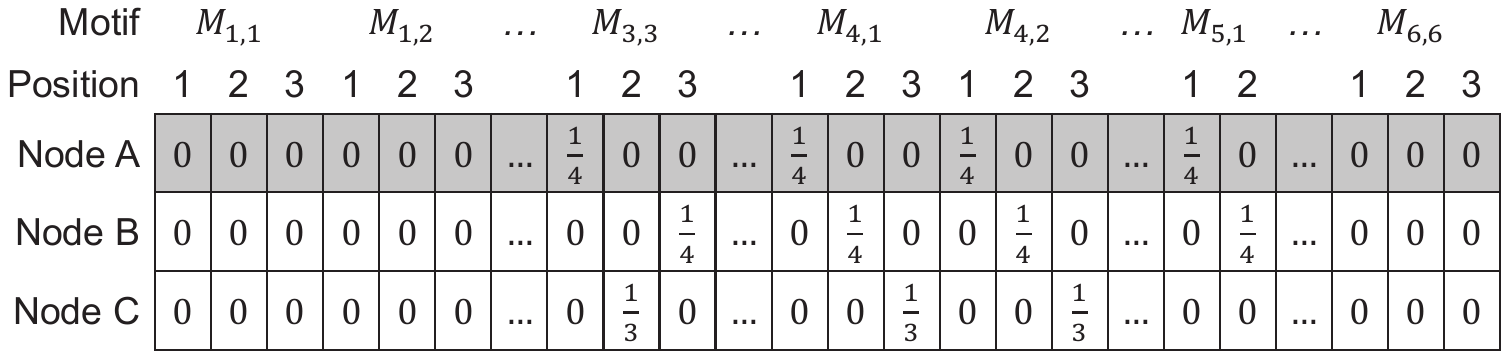}
        \label{fig:tmpp_example_vectors}
    }
    \hfill
    \subfloat[TMPP for node A visualized as heatmaps of temporal motifs and positions]{
        \includegraphics[width=\linewidth]{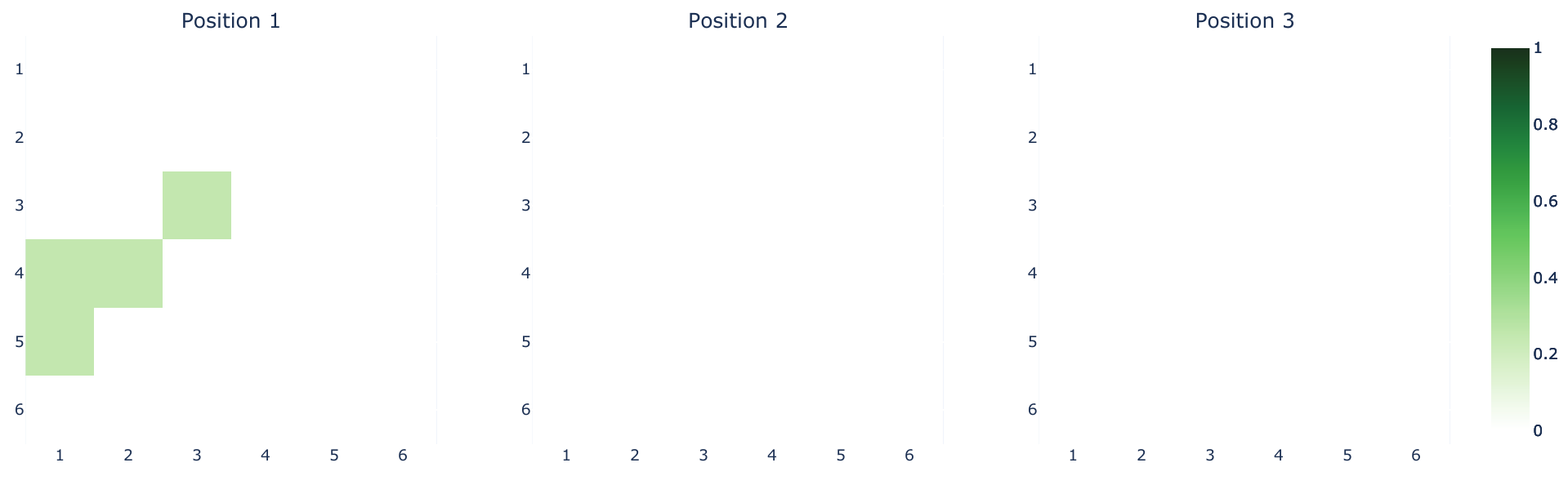}
        \label{fig:tmpp_example_heatmaps}
    }
    \caption[Toy example showing the construction of a temporal motif participation profile (TMPP) for node A from the example network.]{Toy example showing the construction of a temporal motif participation profile (TMPP) for node A from the example network shown in \subref{fig:tmpp_example_graph}.
    The example network contains 4 temporal motifs, and the counts of each of the 3 nodes appearing in the 4 motifs in each position are shown in \subref{fig:tmpp_example_counts}, which are then normalized by the column sums to form TMPP vectors, shown in \subref{fig:tmpp_example_vectors}. 
    Each TMPP vector can be interpreted using heatmaps of temporal motifs and positions, as shown in \subref{fig:tmpp_example_heatmaps} for node A.}
\end{figure}

Consider the toy example in Fig.~\ref{fig:tmpp_example_graph}. The TMPPs for each of the three nodes can be computed by first finding the temporal motifs in the graph. Four 3-edge motifs can be obtained as shown in Fig.~\ref{fig:tmpp_example_graph}.
For example, removing edge 1 results in motif $M_{3,3}$ shown in Fig.~\ref{fig:all_temporal_motifs}, represented by the sequence of edges (1,2), (3,1), (1,3). 
In this motif, node A is in position 1, node C is in position 2, and node B is in position 3. 
The temporal motif position counts for each node are tabulated in Fig.~\ref{fig:tmpp_example_counts}. 
No other temporal motifs are observed, so the counts are 0 for all positions for the other possible temporal motifs.
To obtain the TMPP for a node we normalize its vector of temporal motif counts by its sum, resulting in the vectors shown in Fig.~\ref{fig:tmpp_example_vectors}. 
We can then visualize a TMPP using 3 heatmaps, showing the relative frequency of each temporal motif and position, as shown in Fig.~\ref{fig:tmpp_example_heatmaps}. 
Since a TMPP considers the positions that a node participates in a temporal motif, we also refer to it as a \emph{positioned TMPP}.

\paragraph{Positionless TMPPs}
An alternative to considering node positions in a temporal motif is to simply count the number of times a node participates in a temporal motif, regardless of position. 
Since we consider 36 different temporal motifs, we can also construct a normalized 36-dimensional vector in a similar manner to a TMPP. 
We call this a \emph{positionless TMPP}, as it disregards node position in a temporal motif. 
This is in contrast to the 104-dimensional positioned TMPP we considered earlier, which considers both the temporal motif and position.

\subsection{Clustering and Visualization}
\label{sec:clustering}
A node's TMPP can be viewed as an unsupervised node embedding, where the dimension matches up with the total number of node positions over the types temporal motifs being considered. 
As we showed in Section \ref{sec:tmpp}, this corresponds to $104$ dimensions if we consider all possible 3-edge temporal motifs. 
These TMPPs can be clustered to identify nodes with similar positions in temporal motifs, which correspond to nodes with similar roles in the temporal network.

While many different clustering approaches can be applied to TMPPs, we choose agglomerative hierarchical clustering using Ward's method \cite{ward1963hierarchical} as the linkage criterion. 
This provides us with a more detailed view of how nodes are being grouped together via a dendrogram compared to a flat clustering. 
If a flat clustering is desired, we can then cut the dendrogram at a suitable point based on desired criteria such as interpretability or stability of clusters.

In addition to visualizing TMPPs, we can also create heatmaps for each cluster centroid as a way to visualize the mean TMPP of all nodes in a cluster. 
This provides additional insight into what node behaviors a cluster is capturing and can be compared directly to the heatmap of a node's TMPP, such as the example shown in Fig.~\ref{fig:tmpp_example_heatmaps}.

\section{Experiments}
\label{sec:experiments}

Code to reproduce our experiment results is available at \url{https://github.com/MLNS-Lab/Temporal-Motif-Participation-Profiles}.

\subsection{Simulated Temporal Networks}
We simulate networks from the MULCH generative model \cite{soliman2022multivariate}, which uses structured multivariate Hawkes processes \cite{laub2021elements} to generate temporal networks  that have a variety of different temporal motifs. 
MULCH divides nodes into different blocks of stochastically equivalent nodes with the same role, so we can compare the clusters obtained from clustering the TMPPs computed from the simulated networks to the ground truth block labels to obtain a measure of accuracy. 

\subsubsection{Simulation Scenarios}
We consider two different simulation scenarios. 
In Scenario 1, we choose MULCH parameters to generate primarily the two-node temporal motifs in the bottom left of Fig.~\ref{fig:all_temporal_motifs}. 
Specifically, parameters are chosen so that nodes in both blocks 0 and 1 engage in motif $M_{6,1}$ roughly equally often and in both positions. 
On the other hand, nodes in block 1 are more likely to reciprocate edges from nodes in block 0, so that block 0 generates motifs $M_{5,1}$, $M_{5,2}$, and $M_{6,2}$ frequently in position 1, while block 1 generates them frequently in position 2. 
We simulate networks with $20$ nodes, each randomly assigned to $1$ of $2$ blocks with equal probability. 
(It is easier to recover block labels in larger networks, so choosing a small network adds difficulty to the clustering problem.)

In Scenario 2, we change the MULCH parameters so that two nodes from block 0 often initiate edges with the same single node from block 1, and a node from block 1 often initiates edges with different nodes in block 0. 
This results in nodes in block 0 primarily generating motifs $M_{1,1}$, $M_{1,6}$, and $M_{6,6}$ in positions 1 and 3 (senders) and motifs $M_{4,1}$, $M_{4,3}$, and $M_{6,3}$ in positions 2 and 3 (receivers). 
Nodes in block 1 primarily generate motifs $M_{1,1}$, $M_{1,6}$, and $M_{6,6}$ in position 2 (receiver) and $M_{4,1}$, $M_{4,3}$, and $M_{6,3}$ in position 1 (sender).

\subsubsection{TMPP Clustering Results}
\begin{table}[p]
    \setlength{\tabcolsep}{6pt}
    \centering
    \caption{Clustering accuracy (mean $\pm$ standard error over $100$ simulation runs) in 2 simulated scenarios using the MULCH model with 2 blocks. Highest accuracy is bolded.}
    % The maximum accuracy over both possible permutations of the clusters is reported. 
    % The positioned TMPPs recover the true block structure much more accurately than the positionless ones.}
    \label{tab:sim_accuracy}
    \begin{tabular}{ccc}
    \hline
    Method            & Scenario 1: 2-node motifs & Scenario 2: 3-node motifs \\
    \hline
    Positioned TMPP   & $\mathbf{0.875\pm0.007}$  & $\mathbf{0.933\pm0.008}$       \\
    Positionless TMPP & $0.562\pm0.005$           & $0.725\pm0.008$        \\
    \hline
    \end{tabular}
\end{table}

\begin{figure}[p]
    \centering
    \subfloat[]{
        \includegraphics[height=2.4in]{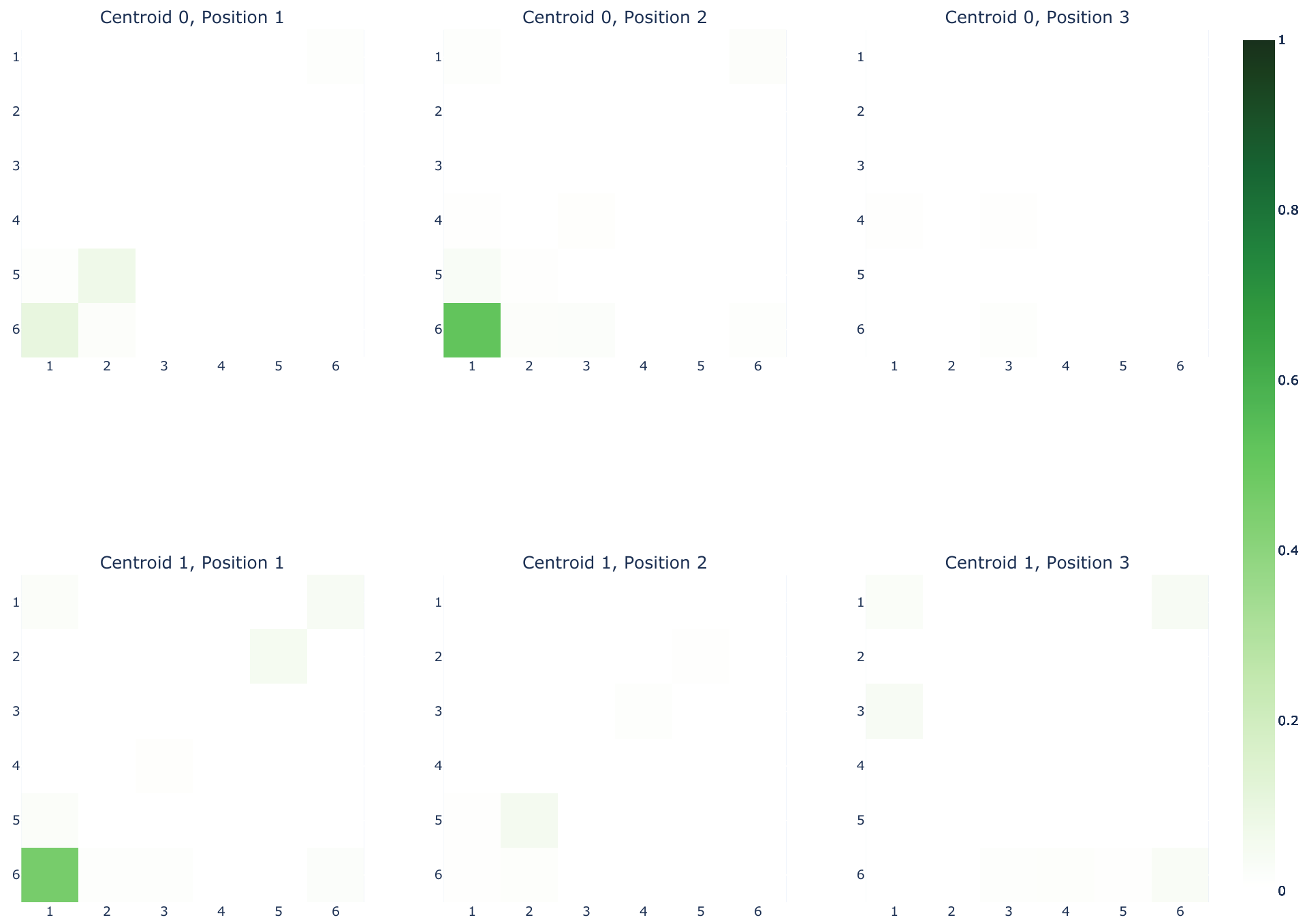}
        \label{fig:sim_scenario_1_tmpp}
    }
    \hfill
    \subfloat[]{
        \includegraphics[height=2.4in]{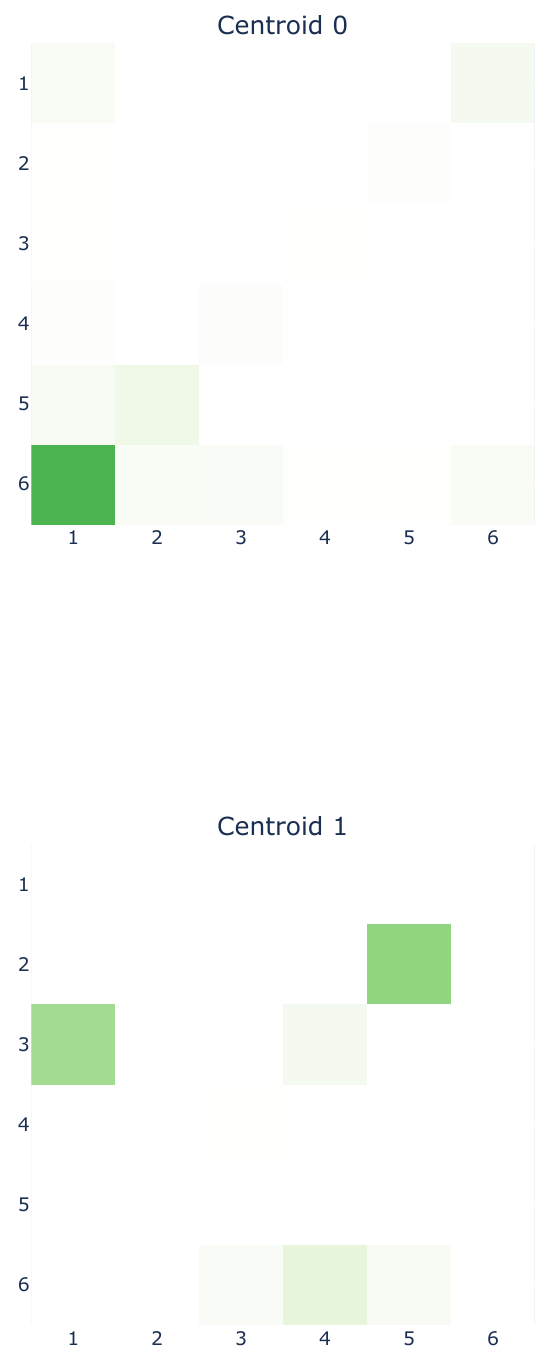}
        \label{fig:sim_scenario_1_positionless}
    }
    \caption[Heatmaps of cluster centroids from Scenario 1]
    {Heatmaps of cluster centroids from Scenario 1 using \subref{fig:sim_scenario_1_tmpp} positioned and \subref{fig:sim_scenario_1_positionless} positionless TMPPs. The positioned TMPP more accurately recovers the true blocks.}
    % The positioned TMPP centroids separate by node position in the 2-node motifs, accurately recovering the true block structure. 
    % The positionless TMPP centroids show that all the nodes that participate in primarily 2-node motifs are placed in cluster 0, with no discernable pattern in cluster 1.}
    \label{fig:sim_scenario_1}
\end{figure}

\begin{figure}[p]
    \centering
    \subfloat[]{
        \includegraphics[height=2.4in]{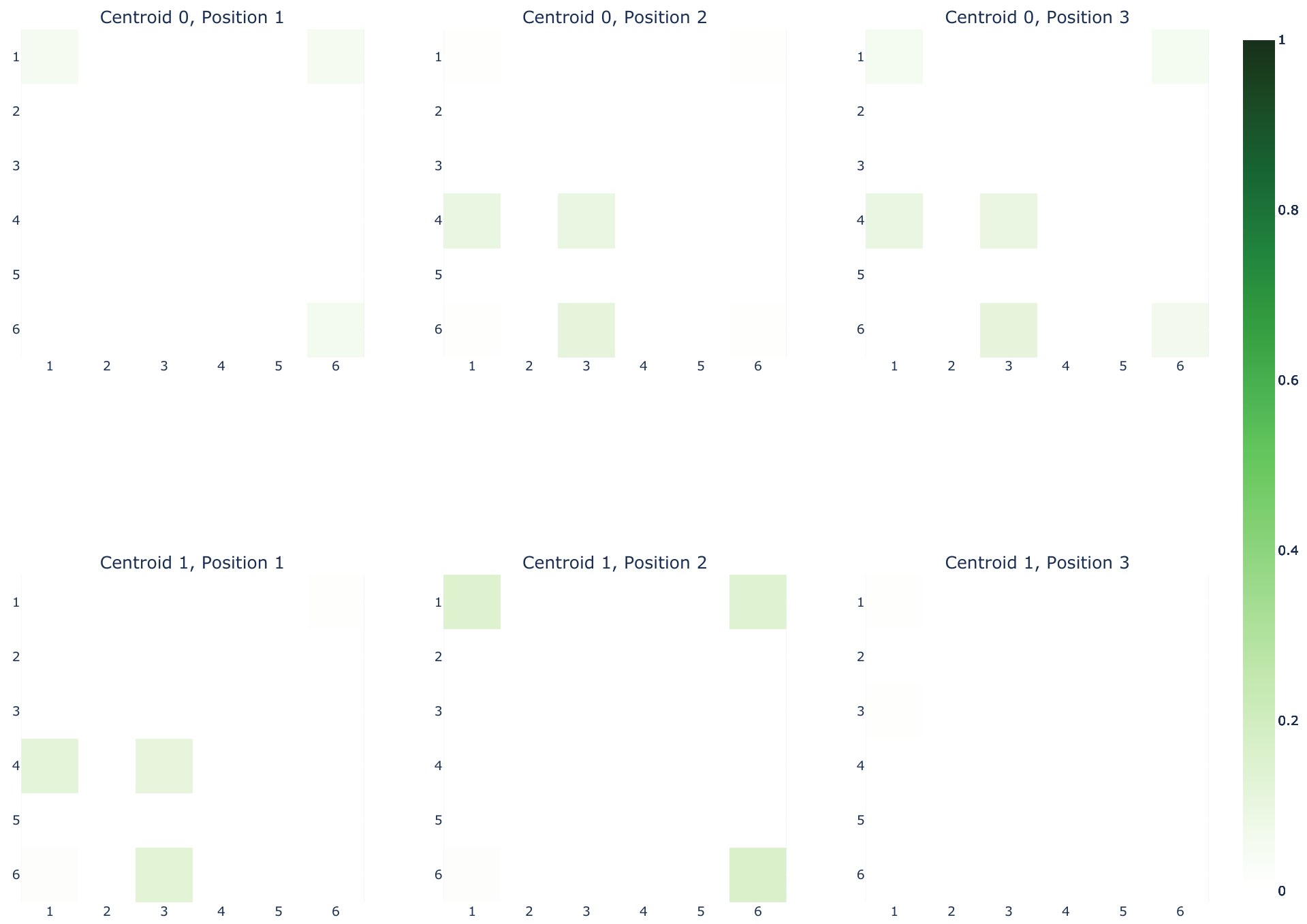}
        \label{fig:sim_scenario_2_tmpp}
    }
    \hfill
    \subfloat[]{
        \includegraphics[height=2.4in]{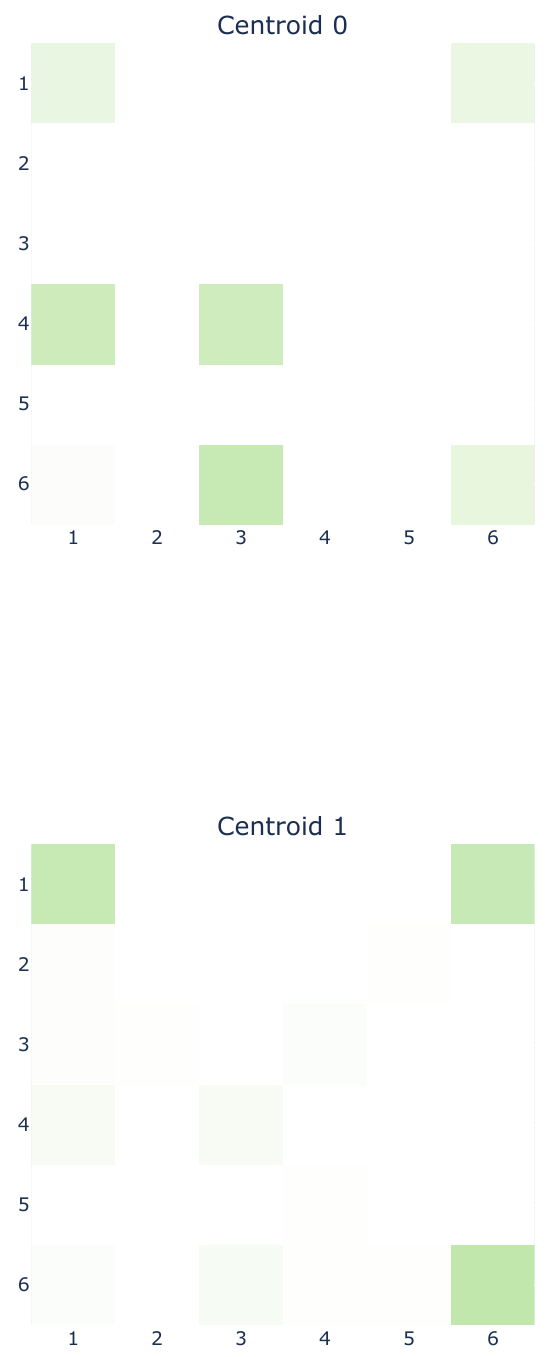}
        \label{fig:sim_scenario_2_positionless}
    }
    \caption[Heatmaps of cluster centroids from Scenario 2]
    {Heatmaps of cluster centroids from Scenario 2 using \subref{fig:sim_scenario_2_tmpp} positioned and \subref{fig:sim_scenario_2_positionless} positionless TMPPs. The positioned TMPP more accurately recovers the true blocks.}
    % The positioned TMPP centroids show clear separation in node positions of the 3-node motifs, accurately recovering the true blocks, while the positionless ones show a mix of the different temporal motifs.}
    \label{fig:sim_scenario_2}
\end{figure}

We perform hierarchical clustering as described in Section \ref{sec:clustering} and flatten into 2 clusters. 
We compute the clustering accuracy by comparing the clusters to the ground truth blocks used to simulate the networks. 
Since clusters are permutation invariant, we consider both possible permutations of the cluster labels and choose the one that gives higher accuracy. 
We compare the clustering accuracy of our (positioned) TMPP with the positionless TMPP that considers only participation in temporal motifs but not node positions.

The results over $100$ simulation runs are shown in Table \ref{tab:sim_accuracy}. 
Notice that, in both scenarios, our proposed TMPP that considers node positions has much higher clustering accuracy. 
The reason for the much higher accuracy is revealed when we consider the TMPPs for the centroids of the clusters, as shown in Figs.~\ref{fig:sim_scenario_1} and \ref{fig:sim_scenario_2}. 
In both cases, the positioned TMPP centroids closely mirror the intended temporal motif generating structure of the simulations, while the positionless centroids either recover clusters with no discernable pattern (centroid 1 in Fig.~\ref{fig:sim_scenario_1_positionless}) or mixed clusters (both centroids in Fig.~\ref{fig:sim_scenario_2_positionless}).
This shows the importance of considering node positions in temporal motifs in order to identify interpretable clusters containing nodes that fulfill similar roles.

\subsection{Case Study: Militarized Interstate Disputes (MIDs)}
We construct a temporal network from the Militarized Interstate Disputes (MIDs) dataset version 5.01 compiled by the Correlates of War project \cite{palmer2022mid5}. 
We use the incident-level data, where each incident represents a threat, display, or use of force that one country directs toward another on a particular day. 
Each country is a node, and each incident is a directed temporal edge from the country that initiates the incident to the country it is directed towards. 
The dataset contains 156 nodes and 5,136 edges and spans from 1992 to 2014. 
Along with the incidents, the dataset also  contains short narrative descriptions of the incidents that are used to code the incident-level data.

We consider all 3-edge temporal network motifs shown in Fig.~\ref{fig:all_temporal_motifs} with a maximum completion time of 7 days, following the analysis of Do and Xu \cite{do2022analyzing} and using their preprocessed data. 
Unlike Do and Xu \cite{do2022analyzing}, we consider only the nodes in the largest strongly connected component, which removes countries that primarily are involved in small, localized conflicts. 
We then remove any nodes that participate in fewer than $10$ temporal motifs, which further removes countries that do not get involved in many interstate disputes. 
We then construct TMPPs and cluster them as described in Sections \ref{sec:tmpp} and \ref{sec:clustering}, respectively.

\subsubsection{TMPP Clustering Results}
\begin{figure}[t]
    \centering
    \includegraphics[width=\linewidth]{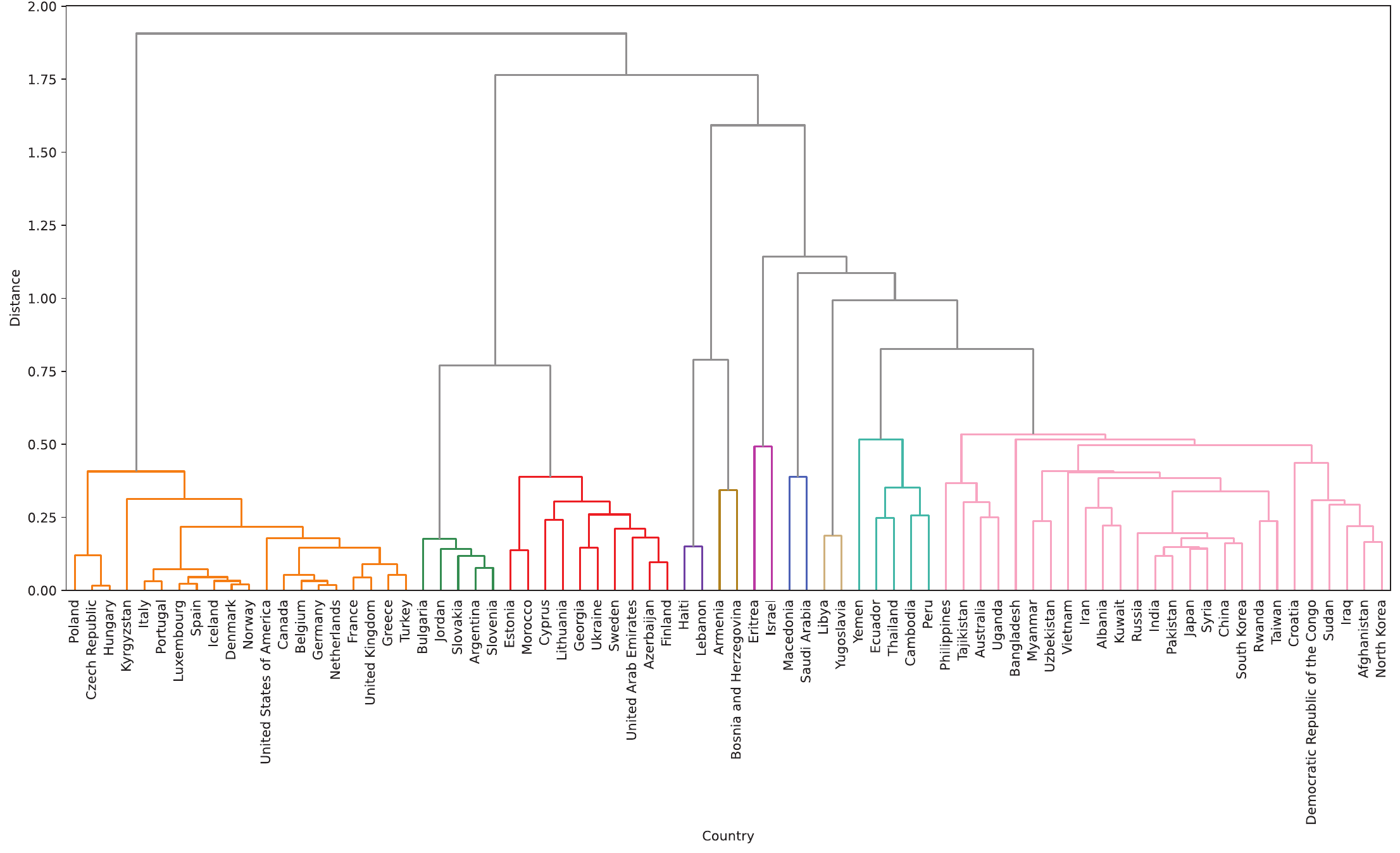}
    \caption{Dendrogram displaying results from hierarchical clustering on the (positioned) TMPPs from the MIDs data. 
    The dendrogram is cut into 10 clusters, as indicated by the colors.
    The orange cluster furthest to the left is denoted as cluster 0, the next cluster in green is denoted as cluster 1, and so on.}
    \label{fig:mids_dendrogram}
\end{figure}

\begin{figure}[tp]
    \centering
    \includegraphics[width=0.95\linewidth]{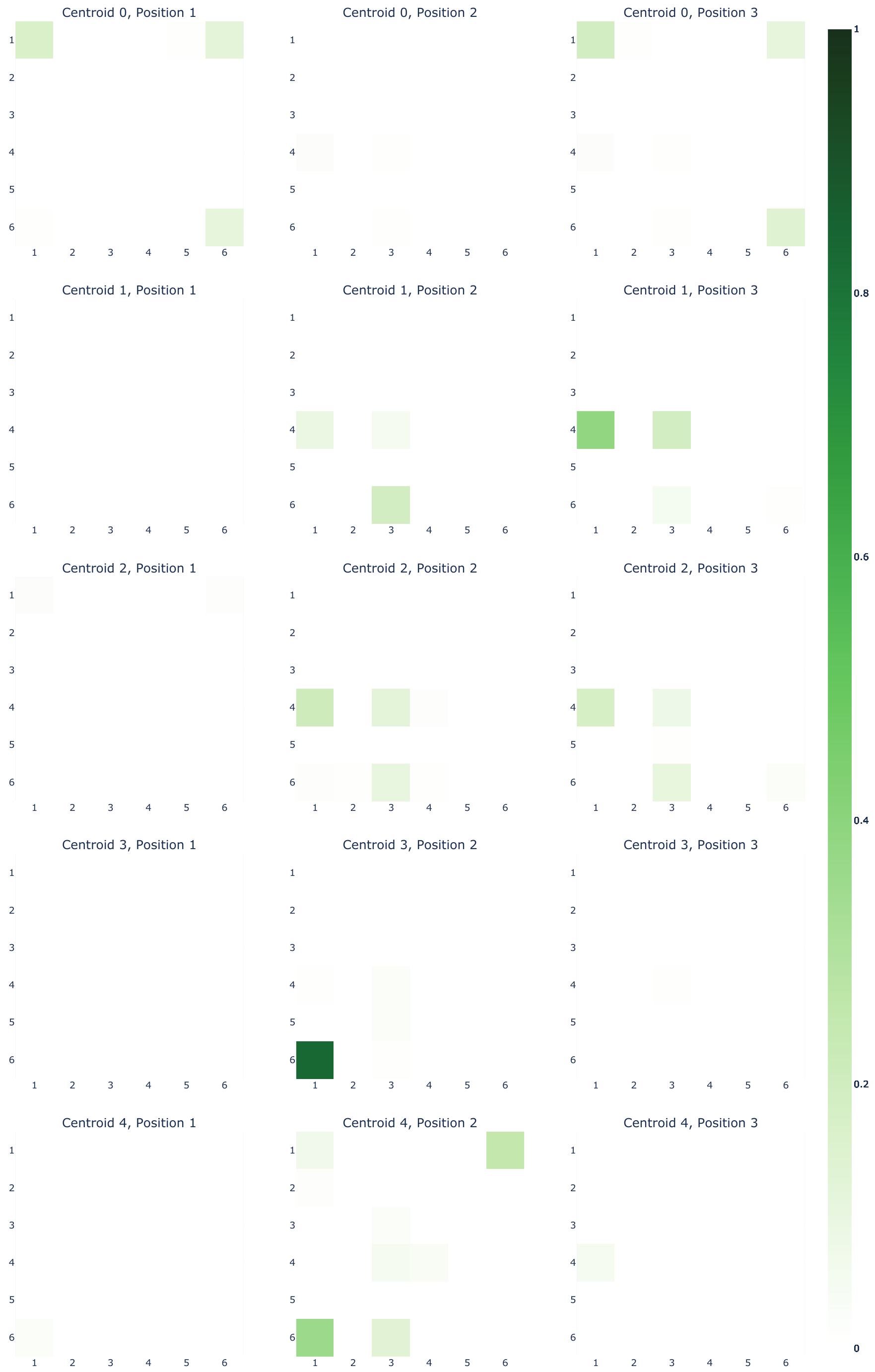}
    \caption{TMPP centroids of the first 5 clusters from the MIDs data. 
    See Fig.~\ref{fig:mids_centroids_2} for the centroids of the last 5 clusters.}
    \label{fig:mids_centroids_1}
\end{figure}

\begin{figure}[tp]
    \centering
    \includegraphics[width=0.95\linewidth]{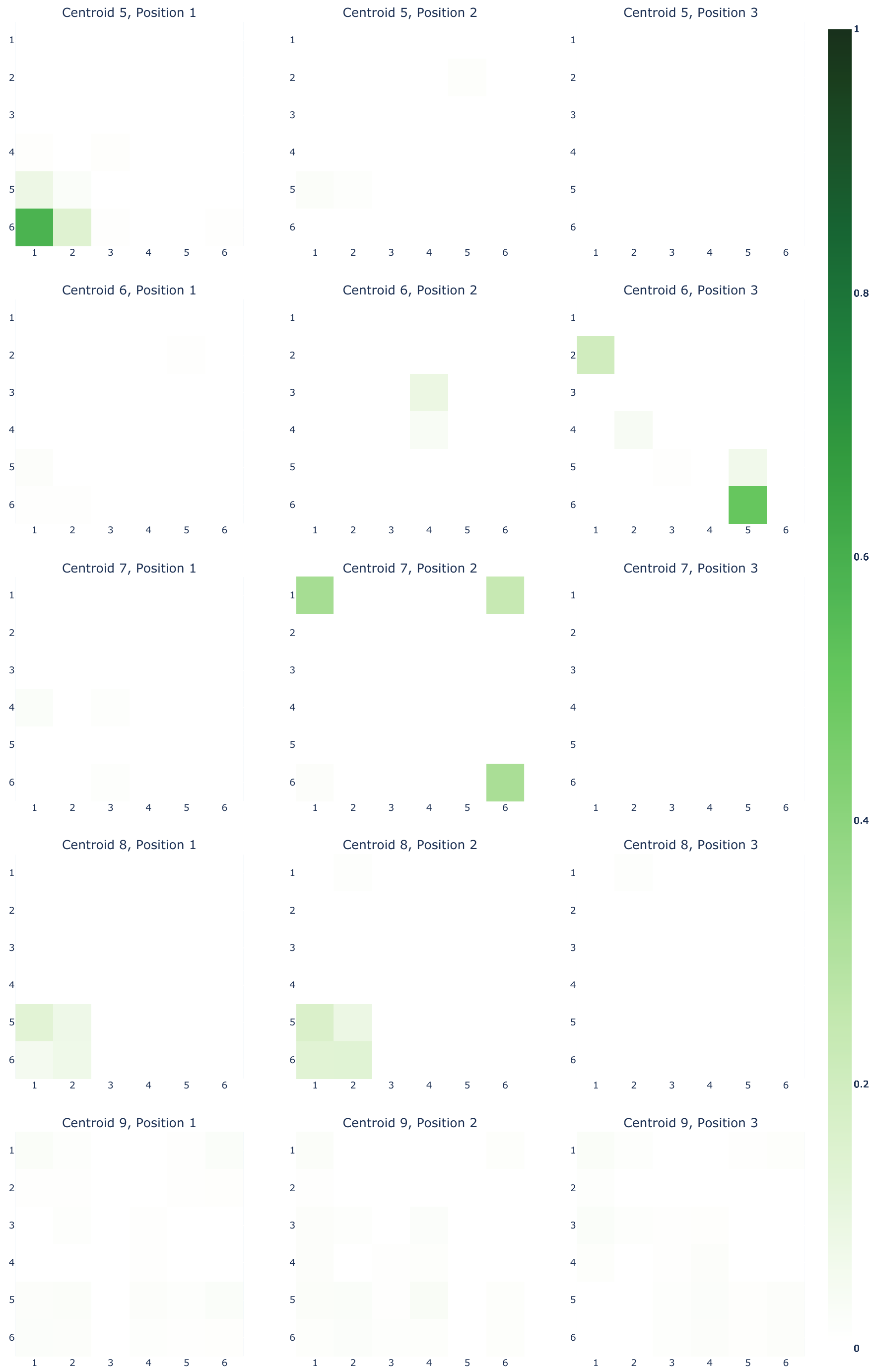}
    \caption{TMPP centroids of last the 5 clusters from the MIDs data. 
    See Fig.~\ref{fig:mids_centroids_1} for the centroids of the first 5 clusters.}
    \label{fig:mids_centroids_2}
\end{figure}

The dendrogram generated from the TMPPs, shown in Fig.~\ref{fig:mids_dendrogram}, indicates the presence of ten clusters. The heatmap representation of the centroids of the ten clusters, which shows the mean behavior of countries in each cluster, are shown in Figs.~\ref{fig:mids_centroids_1} and \ref{fig:mids_centroids_2}. 
Within each centroid, the heatmaps are split into a motif's 3 positions, as shown in Fig.~\ref{fig:tmpp_example_heatmaps}. Within each position, the $6\times 6$ heatmap corresponds to the motifs shown in Fig.~\ref{fig:all_temporal_motifs}. Darker shades of green indicate that countries in that cluster participate in that position of that motif at higher rate.
The following selected cases, from analyzing the MIDs narratives as well as prior work, exemplify how TMPPs can separate and group countries based on their similar roles in MIDs.

\paragraph{Clusters 3 and 5} The MIDs dataset tracks Israel's repeated militarized incidents with Lebanon from 1993 to 2001.  
As noted by Do and Xu \cite{do2022analyzing}, most incidents correspond to Israeli attacks on Hezbollah guerillas in southern Lebanon, with the Lebanese infrequently reciprocating against Israel. As such, Israel is in cluster 5, which displays almost exclusive participation in position 1 of motif $M_{6,1}$ in which a country attacks another 3 times. On the other hand, Lebanon is in cluster 3 which exhibits behavior in position 2 of the $M_{6,1}$ motif, indicating 3 repeated attacks from another country.

Eritrea has been involved in multiple conflicts with Sudan, Djibouti, Yemen, and Ethiopia, often aggressively and with infrequent reciprocation, which explains its presence in cluster 5 with Israel.
Haiti is found in cluster 3 along with Lebanon. 
The United States engaged in threats, shows of force, blockades, and other military action against Haiti without much reciprocation from Haiti.

\paragraph{Clusters 0 and 7} Many NATO countries are grouped in cluster 0, which is characterized by positions 1 and 3 of motifs $M_{1,1}$, $M_{1,6}$, and $M_{6,6}$ in which two countries initiate incidents against a single country.
In 1998 and 1999, NATO engaged in shows of force and airstrikes against Yugoslavia. In 2011, NATO engaged in shows of force, blockades, enforcement of no-fly zones, and bombings against Libya. 
As such, Yugoslavia and Libya are grouped together in cluster 7, which is characterized by the position 2 of motifs $M_{1,1}$, $M_{1,6}$, and $M_{6,6}$, all of which involve being jointly attacked by two countries, mirroring cluster 0.

\paragraph{Cluster 4} Armenia has been involved in a long conflict with Azerbaijan over the Nagorno-Karabakh region, with the occasional Turkish involvement on the side of Azerbaijan. Bosnia and Herzegovina has been the target of Yugoslavia and Croatia in some smaller conflicts. As such, they are both in cluster 4, which has higher activity in position 2 of motifs $M_{6,1}$, $M_{1,6}$, and $M_{6,3}$—indicating that these countries are attacked by one or two other countries.

\paragraph{Cluster 6} Macedonia has been involved in conflicts with Yugoslavia, and Saudi Arabia has been involved in conflicts with Yemen; both Yugoslavia and Yemen have been targets or participants of other conflicts. As a result, Macedonia and Saudi Arabia are grouped together in cluster 6, where countries frequently take on position 3 in motifs $M_{6,5}$, $M_{2,1}$, and $M_{5,5}$ and position 2 in motif $M_{3,4}$. This indicates that these countries are attacked by a country that was or will be involved in other conflicts.

\begin{figure}[t]
    \centering
    \includegraphics[width=\linewidth]{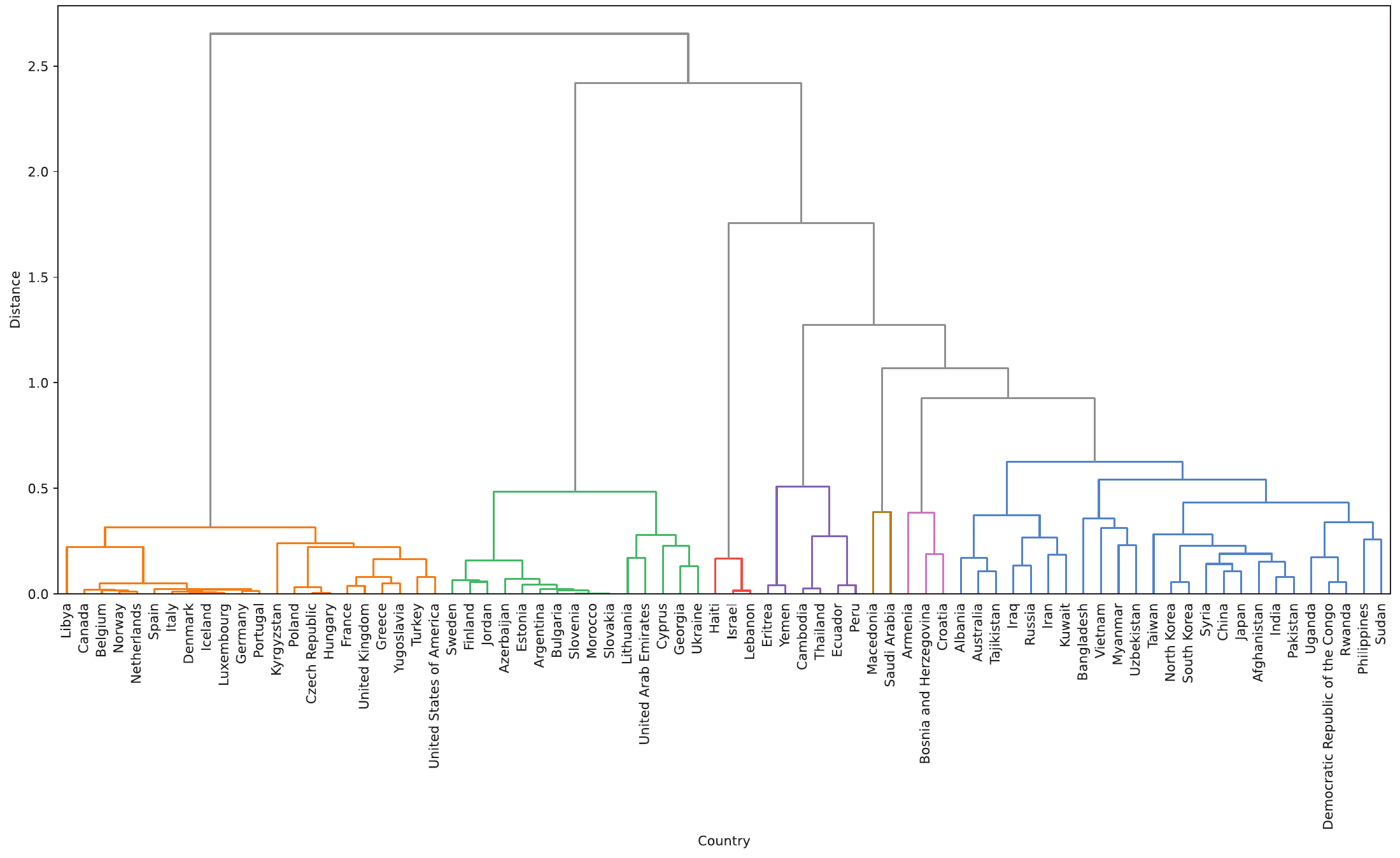}
    \caption{Dendrogram displaying results from hierarchical clustering on the positionless TMPPs from the MIDs data. 
    The dendrogram is cut into 7 clusters, as indicated by the colors.
    The orange cluster furthest to the left is denoted as cluster 0, the next cluster in green is denoted as cluster 1, and so on.    
    }
    \label{fig:mids_dendrogram_positionless}
\end{figure}

\subsubsection{Comparison to Positionless TMPPs}
To illustrate the importance of accounting for a node's position in a temporal motif, we provide a comparison to TMPPs without node positions.
The hierarchical clustering results for these positionless TMPPs is shown in Fig.~\ref{fig:mids_dendrogram_positionless}. 
From this dendrogram, we find that nodes that tend to engage in the same motifs tend to be grouped together, rather than nodes that occupy the same roles in motifs. 
We highlight some specific cases in the following.

\paragraph{Israel and Lebanon}
One of the first merges made by the hierarchical clustering algorithm on the positionless TMPPs groups Israel and Lebanon together in  cluster 0. 
As we discussed earlier, Israel and Lebanon have many instances of motif $M_{6,1}$, corresponding to Israel initiating incidents towards Lebanon three times. 
Thus, Israel and Lebanon occupy opposite roles in the motifs, which is not accounted for by the positionless TMPPs.  
When we compare with the positioned TMPPs, Israel and Lebanon are in very different clusters in the TMPPs shown in Fig.~\ref{fig:mids_dendrogram}, forming the third to last merge in the dendrogram. 

\paragraph{Yugoslavia, Libya, and NATO}
In both the positioned and positionless TMPPs, we find that cluster 0 contains many NATO members (colored orange in both Figs.~\ref{fig:mids_dendrogram} and \ref{fig:mids_dendrogram_positionless}). 
When using the positioned TMPPs, Yugoslavia and Libya are found in cluster 7, denoting that the NATO members were initiating incidents towards Yugoslavia and Libya. 
On the other hand, the positionless TMPPs place Yugoslavia and Libya in cluster 0 with the NATO members because they co-occur in the same motifs, but in differing roles. 
Thus, the positioned TMPPs provide a role-based embedding and clustering of nodes, while the positionless TMPPs are only capturing co-occurrence of nodes in motifs in this instance. 

\section{Conclusion}
We proposed the temporal motif participation profile (TMPP) as a way of capturing how nodes behave in temporal networks. 
The TMPP provides a human-interpretable unsupervised node embedding that can be visualized in the form of heatmaps denoting frequencies at which nodes participate in different positions of temporal motifs. 
We showed that nodes with similar TMPPs often shared similar roles within a temporal network, and clustering TMPPs could identify such nodes. 
The inclusion of node positions in temporal motifs was crucial to capture this role similarity, as we found that participation in temporal motifs in any position, which we called a positionless TMPP, tended to cluster together nodes that participated in the same edges of a temporal motif rather than roles. 
We illustrated the latter on a case study involving militarized interstate disputes.

\paragraph{Limitations and Future Work}
In this paper, we considered all possible 3-edge temporal motifs; however, our proposed TMPPs can be applied to any types of temporal motifs, enabling the analysis of larger motifs or specific motifs of interest. 
TMPPs can also be applied to other definitions of temporal motifs \cite{sariyuce2025powerful}. 
While our focus was on TMPPs as an analysis tool and not on scalability, we note that there has been research on scaling up methods for counting temporal motifs, including sampling-based methods \cite{liu2019sampling}. 
One could potentially create similar approaches to compute approximate TMPPs at scale to analyze extremely large temporal networks, which would be an interesting avenue for future work.

\begin{credits}
\subsubsection{\ackname}
We thank Hung Do for insightful discussions about this work. 
This material is based upon work supported by the National Science Foundation grant IIS-2318751. 

\subsubsection{\discintname}
The authors have no competing interests to declare that are
relevant to the content of this article.
\end{credits}
\bibliographystyle{splncs04}
\bibliography{references}
\end{document}